\documentclass[journal]{IEEEtran}

\usepackage{cite}
\usepackage{url}
\usepackage{graphicx}
\usepackage{color}
\usepackage{bm}
\usepackage{cite}
\usepackage{amsmath,amssymb,amsfonts}
\usepackage{algorithmic}
\usepackage{graphicx}
\usepackage{textcomp}
\usepackage{algorithmic}
\usepackage{algorithm}
\usepackage{array}
\usepackage[caption=false,font=normalsize,labelfont=sf,textfont=sf]{subfig}
\usepackage{stfloats}
\usepackage{url}
\usepackage{verbatim}
\usepackage{enumerate}
\usepackage{tikz}
\usepackage{bm}
\usepackage{threeparttable,booktabs}
\usepackage{blkarray}
\usepackage{amsmath}
\usepackage{multirow}

\usepackage{amsmath}
\usepackage{amsthm,amsfonts}
\usepackage{pgfplots}

\usepackage{tikz}
\usepackage[customcolors]{hf-tikz}
\usepgfplotslibrary{colorbrewer}
\usetikzlibrary{%
  3d,
  arrows,%
  shapes.misc,
  shapes.arrows,%
  chains,%
  matrix,%
	fit,%
  positioning,
  scopes,%
  decorations.pathmorphing,
  shadows,%
  decorations.markings,
  shapes.geometric,
  arrows.meta,bending,
	patterns,
	intersections, 
	pgfplots.fillbetween
}

\usepackage{tikz-3dplot}
\usepackage[pdfusetitle]{hyperref}
\hypersetup{
    colorlinks,
    linkcolor={red!50!black}, 
    citecolor={blue!50!black},
    urlcolor={blue!50!black}
}

\usepackage{cellspace}
\providecommand*{\mat}[1]{\mathbf#1}

\DeclareMathAccent{\ring}{\mathalpha}{operators}{"17}

\newcommand{\ie}{\textit{i.e.}\/, }
\newcommand{\eg}{\textit{e.g.}\/, }

\colorlet{dpurple}{blue!50!red}
\colorlet{dblue}{blue!50!black}
\colorlet{dgreen}{green!50!black}
\colorlet{dred}{red!50!black}
\colorlet{dyellow}{yellow!50!black}
\colorlet{dorange}{orange!50!black}
\definecolor{metal}{RGB}{218,165,32}
\definecolor{diel}{RGB}{1,165,32}
\definecolor{antenna}{RGB}{100,150,162}
\definecolor{breg}{rgb}{0.2,0.6,0.8}%
\definecolor{preg}{rgb}{0.8,0.2,0.2}%
\definecolor{reg}{RGB}{218,165,32}

\graphicspath{{figures/}{tikz/}}

\makeatother

\begin{document}
\title{Generalized Scattering Matrix Synthesis: Independent Region Decomposition for Hybrid Antenna--Scatterer Systems}

\author{Chenbo Shi, Shichen Liang, Jin Pan, Xin Gu and Le Zuo
\thanks{Manuscript received Nov. 19, 2024. Revised Jun. 18, 2025. Accepted Aug.~6, 2025. (\textit{Corresponding author: Jin Pan.})}
\thanks{Chenbo Shi, Shichen Liang, Jin Pan and Xin Gu are with the School of Electronic Science and Engineering, University of Electronic Science and Technology of China, Chengdu 611731 China  (e-mail: chenbo\_shi@163.com; lscstu001@163.com; panjin@uestc.edu.cn; xin\_gu04@163.com).}
\thanks{Le Zuo is with The 29th Research Institute of China Electronics Technology Group Corporation (e-mail: zorro1204@163.com)}
}


\maketitle

\begin{abstract}
This paper presents a unified formulation for synthesizing the generalized scattering matrix (GS-matrix) of hybrid electromagnetic systems comprising arbitrary numbers of antennas and scatterers. The proposed method provides a modular region decomposition framework that enables efficient analysis of electromagnetic interactions between distinct structures, under the relaxed geometric condition that the constituents are separable by a plane. By leveraging the addition theorem of vector spherical wavefunctions (VSWFs), a compact matrix representation is derived to assemble the GS- and S-matrices of individual components into the overall system response. This formulation generalizes and extends prior methods developed for either multiple scattering or antenna array analysis, and is particularly suited to configurations where substructures may be repositioned or reused. Numerical examples are provided to validate the accuracy and versatility of the method, including scenarios involving tightly spaced components and rotational variations in substructure layout.
\end{abstract}


\section{Introduction}

\IEEEPARstart{T}{he} scattering matrix (S-matrix) is a fundamental paradigm for characterizing how a scatterer interacts with and modulates an incident electromagnetic field. Typically, the S-matrix is defined based on an expansion of the electromagnetic field using vector spherical wave functions (VSWFs) \cite{ref_scattering_Montgomery,ref_Smat_ori}. Leveraging the advantageous translational properties of VSWFs \cite{ref_sph_addition1,ref_sph_addition2,ref_Kristensson_booklet}, the S-matrix method enables analytical modeling of mutual coupling among multiple structures, and has been widely applied to the analysis of multiple scattering problems \cite{ref_Multiple1,ref_Multiple2,ref_Multiple3,ref_Multiple_E3}. Moreover, since the eigenvalues of the S-matrix are directly related to the system’s characteristic modes, it also serves as an efficient tool for modal decomposition of multi-object systems \cite{ref_mysyn_CMA}.

The S-matrix method exhibits key features of region decomposition: the modeling of individual structures is independent of their mutual coupling, allowing complex multi-object systems to be partitioned into simpler, modular, and independently analyzable subproblems, whose results can then be assembled to yield the overall system response. Compared to conventional domain decomposition techniques \cite{ref_DDM_FEM1,ref_DDM_FEM2,ref_DDM_FDM1,ref_DDM_FDM2,ref_DDM_MoM1}, its major advantage lies in representing multiple scattering as a series of explicit matrix operations, without requiring iterative solutions of coupled boundary problems. This non-iterative formulation enables mutual coupling to be evaluated entirely in a post-processing stage, facilitating efficient reuse of precomputed substructure responses when exploring different spatial arrangements or modifying only part of the system. This modularity significantly reduces redundant computations and supports scalable design workflows. Furthermore, with customized acceleration techniques \cite{ref_Fast_Tmat}, the method can achieve a computational complexity as low as $\mathcal{O}(N\log N)$ in certain scenarios, exhibiting excellent scalability. A classical limitation, however, is that the minimal enclosing spheres of different structures must remain non-overlapping to ensure the convergence of the spherical wave expansion and the validity of analytical translation formulas.

An important extension of this framework is the generalized scattering matrix (GS-matrix), which incorporates antenna port excitations into the S-matrix formulation \cite{ref_scattering_Montgomery,ref_sph_near_measure}. The GS-matrix has been widely adopted to evaluate mutual coupling among multiple antennas, including in applications such as spherical near-field measurements \cite{ref_sph_near_measure} and array performance prediction \cite{ref_3D_FEM}. A recent study \cite{ref_myGSM} reported further advancements in GS-matrix-based techniques. In addition, the GS-matrix has also been combined with the S-matrix of nearby scatterers \cite{ref_Egypt} to evaluate antenna performance in the presence of surrounding structures, such as reflector antennas. Although the method in \cite{ref_Egypt} demonstrates the capability of the S-matrix framework for handling single antenna-scatterer interactions, it has not yet been generalized to systems involving arbitrary numbers of antennas and scatterers.

This paper aims to establish a general GS-matrix formulation for hybrid systems comprising arbitrary antennas and scatterers, while relaxing the geometric constraints imposed by existing methods. To this end, we employ plane vector wave functions to assist in the translation of VSWFs \cite{ref_translation_PRA}. The formulas from \cite{ref_translation_PRA} are reformulated to support near-field translation within a real-valued VSWF framework. This strategy not only unifies previous results for purely antenna-based systems (\eg \cite{ref_3D_FEM,ref_myGSM}), but also incorporates classical multiple scattering models \cite{ref_Multiple1,ref_Multiple2,ref_Multiple3,ref_Multiple_E3}, and extends the applicability of the scattering matrix paradigm to configurations where structures are merely required to be separable by a plane, rather than having non-overlapping circumscribing spheres.

Finally, several numerical examples are presented to validate the accuracy, modularity, and applicability of the proposed method. The results demonstrate that, compared to existing approaches, the proposed framework significantly broadens the usable scope of the S-matrix method. While further investigation is required to fully address scenarios involving electrically connected or conformal structures, the current findings indicate strong adaptability and extensibility.

\section{A Review of Generalized Scattering Matrix}
\label{SecII}

\begin{figure}[!t]
  \centering
  \includegraphics[]{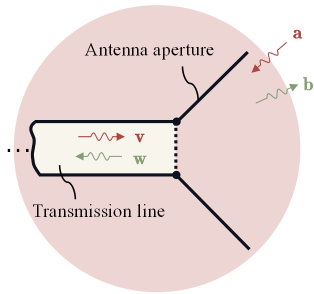}
  \caption{A typical electromagnetic system in which the incoming and outgoing states include spatially radiated waves and guided waves propagating in transmission lines.}
  \label{f_Rad_ScaModel}
\end{figure}

Figure \ref{f_Rad_ScaModel} schematically illustrates the standard scenario captured by the GS-matrix. Incoming and outgoing states encompass VSWFs external to the structure (represented by expansion vectors $\mathbf{a}$ and $\mathbf{b}$), as well as eigenmodes propagating within waveguide transmission lines (represented by vectors $\mathbf{v}$ and $\mathbf{w}$, respectively). Consistent with the notation established in \cite{ref_myGSM}, the GS-matrix is expressed as 
\begin{equation}
  \label{eq18}
  \tilde{\mathbf{S}}=\begin{bmatrix}
    \mathbf{\Gamma }&		\mathbf{R}\\
    \mathbf{T}&		\mathbf{S}
  \end{bmatrix}
\end{equation}
and
\begin{equation}
  \begin{bmatrix}
    \mathbf{w}\\
    \mathbf{b}\\
  \end{bmatrix}=\begin{bmatrix}
    \mathbf{\Gamma }&		\mathbf{R}\\
    \mathbf{T}&		\mathbf{S}
  \end{bmatrix}\begin{bmatrix}
    \mathbf{v}\\
    \mathbf{a}\\
  \end{bmatrix}.
\end{equation}
Each sub-block of $\tilde{\mathbf{S}}$ carries clear physical interpretation: $\mathbf{\Gamma}$ denotes the port scattering matrix, $\mathbf{R}$ is the receiving matrix, $\mathbf{T}$ the transmitting matrix, and $\mathbf{S}$ the conventional scattering matrix (S-matrix). To avoid ambiguity, we refer to $\mathbf{\Gamma}$ explicitly as the ``S-parameters," distinguishing them clearly from the S-matrix (noting that, for single-port and single-mode configurations, $\mathbf{\Gamma}$ simplifies to a scalar reflection coefficient).

In practical applications, a commonly adopted variant of the GS-matrix formulation omits the intrinsic transformation between incoming and outgoing states that is independent of the structural properties:
\begin{equation}
  \label{eq3}
  \begin{bmatrix}
    \mathbf{w}\\
    \mathbf{f}\\
  \end{bmatrix}=\begin{bmatrix}
    \mathbf{\Gamma }&		\mathbf{R}\\
    \mathbf{T}&		\mathbf{S-1}
  \end{bmatrix}\begin{bmatrix}
    \mathbf{v}\\
    \mathbf{a}\\
  \end{bmatrix}
\end{equation}
where $\mathbf{a}$ denotes the incident wave expansion vector, and $\mathbf{f} = \mathbf{b} - \mathbf{a}$ represents the scattered wave expansion vector. The identity matrix is denoted by $\mat{1}$. This modified formulation facilitates mutual coupling analysis between distinct structures and serves as the foundational expression throughout this paper.

Two specific degenerate forms of \eqref{eq3} are particularly noteworthy. For a purely scattering structure without waveguide ports, only the lower-right subblock, $\mathbf{S}-\mathbf{1}$, remains, reducing the relation to
\begin{equation}
  \mathbf{f} = (\mathbf{S} - \mathbf{1}) \mathbf{a}.
\end{equation}
Conversely, for microwave circuits such as waveguides, which involve no radiation into free space, only the upper-left subblock $\mathbf{\Gamma}$ is relevant, yielding the classical port-based scattering relation: $\mathbf{w} = \mathbf{\Gamma} \mathbf{v}$.

The dimensions of $\mathbf{S}$ and $\mathbf{\Gamma}$ are $j \times j$ and $e \times e$, respectively, where $j$ denotes the number of VSWFs, and $e$ represents the total number of propagating eigenmodes in the waveguide ports. The value of $j$ is determined by the truncation degree $L_{\max}$ via $j = 2L_{\max}(L_{\max} + 2)$, where $L_{\max}$ is typically selected using the empirical formula \cite{ref_Sph_deg_trunction}:
\begin{equation}
  L_{\max} = \left\lceil k R_{\min} + 7\sqrt[3]{k R_{\min}} + 3 \right\rceil,
\end{equation}
with $R_{\min}$ denoting the minimal radius of the circumscribing sphere of the structure.

This paper does not address the computation of the GS-matrix or S-matrix for individual antennas or scatterers, and assumes that these matrices are precomputed using, for example, the methods described in \cite{ref_3D_FEM,ref_myGSM}. The focus here is on assembling the overall GS- or S-matrix of the composite system based on these known submatrices. In this context, \cite{ref_myGSM} serves as practical means for generating the GS-matrix of individual antenna elements, which are then incorporated into the modular synthesis framework presented in this work.

\section{Synthesis of the Generalized Scattering Matrix for Antenna-Scatterer Hybrid System}
\label{Sec_III}

\begin{figure}[!t]
  \centering
  \includegraphics[]{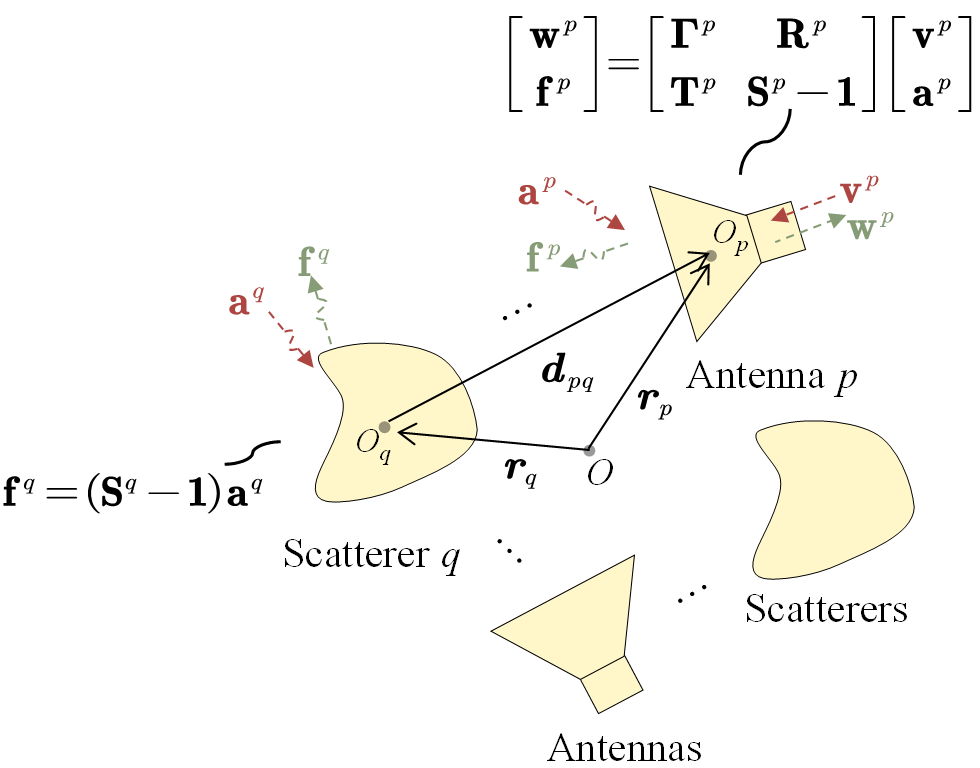} 
  \caption{Illustration of a hybrid electromagnetic system consisting of multiple antennas and scatterers.}
  \label{f_HybridSystem}
\end{figure}

We now consider a general hybrid electromagnetic system composed of $M$ antennas and $N$ scatterers, thus comprising a total of $M + N$ distinct structures, labeled sequentially as $p = 1, 2, \dots, M, M+1, \dots, M+N$, as illustrated in Fig.~\ref{f_HybridSystem}. For convenience, we define the set $A = {1, 2, \dots, M}$ representing antennas and the set $S = {M+1, M+2, \dots, M+N}$ representing scatterers, hence $p \in A \cup S$. Each structure is described within its own local coordinate system, \textit{assumed to differ from the global coordinates only by translational displacement}.

The electromagnetic characteristics of each individual structure are fully encapsulated within its local GS- or S-matrix. Specifically, the electromagnetic interactions of the $p$-th structure can be expressed through its corresponding local matrices as follows:
\begin{equation}
  \begin{cases}
  \label{eq6}
  \mathbf{\Gamma }^p\mathbf{v}^p+\mathbf{R}^p\mathbf{a}^p=\mathbf{w}^p\\
    \mathbf{T}^p\mathbf{v}^p+(\mathbf{S}^p-\mathbf{1})\mathbf{a}^p=\mathbf{f}^p\\
  \end{cases},p\in A
\end{equation}
for antenna-type structures with ports, and 
\begin{equation}
  \label{eq7}
  (\mathbf{S}^p-\mathbf{1})\mathbf{a}^p=\mathbf{f}^p,\  p \in S.
\end{equation}
for scatterer-type structures without waveguide ports.

The incident wave expansion vector $\mathbf{a}^p$ at structure $p$ consists of two distinct contributions: the externally imposed direct incident field $\mathbf{a}_\mathrm{d}^p$ (e.g., plane wave illumination), and the multiple scattered fields $\mathbf{f}^q$ arising from other structures ($q\ne p$). Consequently, the total incident field at structure $p$ can be expressed as
\begin{equation}
  \label{eq8}
  \mathbf{a}^p=\mathbf{a}^p_\mathrm{d}+\sum_{q\ne p}{\bm{\mathcal{G}} _{pq}}\mathbf{f}^q
\end{equation}
where $\bm{\mathcal{G}} _{pq}=\bm{\mathcal{G}}(k\boldsymbol{d}_{pq})$ is the translation matrix responsible for transforming scattered fields from the local coordinate system of structure $q$ into incident fields expressed in the local coordinate system of structure $p$. Details regarding the construction and properties of these translation matrices are elaborated in Sec.~\ref{Sec_IV}. $k$ represents the free-space wavenumber. Note that in \eqref{eq8}, the index $q$ spans all elements within the combined set $A \cup S$, and the equation applies to all structures $p$.

Substituting \eqref{eq8} into \eqref{eq6} and \eqref{eq7} results in:
\begin{equation}
  \begin{cases}
    \mathbf{\Gamma }^p\mathbf{v}^p+\mathbf{R}^p\mathbf{a}^p_\mathrm{d}+\mathbf{R}^p \displaystyle\sum_{q\ne p}{\bm{\mathcal{G}}_{pq}\mathbf{f}^q}=\mathbf{w}^p\\
    \mathbf{T}^p\mathbf{v}^p+(\mathbf{S}^p-\mathbf{1})\mathbf{a}^p_\mathrm{d}+(\mathbf{S}^p-\mathbf{1}) \displaystyle\sum_{q\ne p}{\bm{\mathcal{G}}_{pq}\mathbf{f}^q}=\mathbf{f}^p
  \end{cases}
\end{equation}
for antennas ($p\in A$), and 
\begin{equation}
  \left( \mathbf{S}^p-\mathbf{1} \right) \mathbf{a}^p_\mathrm{d}+\left( \mathbf{S}^p-\mathbf{1} \right) \sum_{q\ne p}{\bm{\mathcal{G}}_{pq}\mathbf{f}^q}=\mathbf{f}^p
\end{equation}
for scatterers ($p\in S$). Although iterating the index $p$ across the complete set $A \cup S$ yields a fully determined system of equations, solving these equations directly can become cumbersome. Hence, it is advantageous to rewrite them in a more compact matrix representation as follows: 
\begin{equation}
  \label{eq11}
  \begin{cases}
    \hat{\mathbf{\Gamma}}^A\hat{\mathbf{v}}^A+\hat{\mathbf{R}}^A\hat{\mathbf{a}}_\mathrm{d}^{A}+\hat{\mathbf{R}}^A \bm{\mathcal{G}}^A\hat{\mathbf{f}}=\hat{\mathbf{w}}^A\\
    \hat{\mathbf{T}}^A\hat{\mathbf{v}}^A+\left( \hat{\mathbf{S}}^A-\mathbf{1} \right) \hat{\mathbf{a}}_\mathrm{d}^{A}+\left( \hat{\mathbf{S}}^A-\mathbf{1} \right) \hat{\bm{\mathcal{G}}}^A\hat{\mathbf{f}}=\hat{\mathbf{f}}^A\\
    \left( \hat{\mathbf{S}}^S-\mathbf{1} \right) \hat{\mathbf{a}}_\mathrm{d}^{S}+\left( \hat{\mathbf{S}}^S-\mathbf{1} \right) \hat{\bm{\mathcal{G}}}^S\hat{\mathbf{f}}=\hat{\mathbf{f}}^S.\\
  \end{cases} 
\end{equation}

In this notation, matrices with superscripts $A$ or $S$ represent blocks corresponding explicitly to antennas or scatterers, respectively. These block-diagonal matrices are explicitly defined as:
\begin{equation}
  \begin{split}
  &\hat{\mathbf{\Gamma}}^A=\mathrm{diag}\left( \mathbf{\Gamma }^1, \cdots ,\mathbf{\Gamma }^M \right) ,\hat{\mathbf{T}}^A=\mathrm{diag}\left( \mathbf{T}^1,\cdots, \mathbf{T}^M \right),\\
  &\hat{\mathbf{R}}^A=\mathrm{diag}\left( \mathbf{R}^1,\cdots, \mathbf{R}^M \right),\hat{\mathbf{S}}^A=\mathrm{diag}\left( \mathbf{S}^1,\cdots \mathbf{S}^M \right) ,\\
  &\hat{\mathbf{S}}^S=\mathrm{diag}\left( \mathbf{S}^{M+1},\cdots, \mathbf{S}^{M+N} \right) 
  \end{split}
\end{equation}
and the associated stacked vectors are expressed as
\begin{equation}
  \begin{split}
    &\hat{\mathbf{v}}^A=\begin{bmatrix}
      \mathbf{v}^1\\
      \vdots\\
      \mathbf{v}^M\\
    \end{bmatrix} ,\hat{\mathbf{w}}^A=\begin{bmatrix}
      \mathbf{w}^1\\
      \vdots\\
      \mathbf{w}^M\\
    \end{bmatrix} , \hat{\mathbf{a}}_\mathrm{d}^{A}=\begin{bmatrix}
      \mathbf{a}_\mathrm{d}^{1}\\
      \vdots\\
      \mathbf{a}_\mathrm{d}^{M}\\
    \end{bmatrix} ,
    \\
    &\hat{\mathbf{a}}_\mathrm{d}^{S}=\begin{bmatrix}
      \mathbf{a}_\mathrm{d}^{M+1}\\
      \vdots\\
      \mathbf{a}_\mathrm{d}^{M+N}\\
    \end{bmatrix}  ,\hat{\mathbf{f}}^A= \begin{bmatrix}
      \mathbf{f}^1\\
      \vdots\\
      \mathbf{f}^M\\
    \end{bmatrix} ,\hat{\mathbf{f}}^S= \begin{bmatrix}
      \mathbf{f}^{M+1}\\
      \vdots\\
      \mathbf{f}^{M+N}\\
    \end{bmatrix}.
  \end{split}
\end{equation}
Additionally, we define the aggregated scattered-field vector as: $\hat{\mathbf{f}}=\begin{bmatrix}
  \hat{\mathbf{f}}^A\\
  \hat{\mathbf{f}}^S
\end{bmatrix}$. The translation matrices $\hat{\bm{\mathcal{G}}}^{A}$ and $\hat{\bm{\mathcal{G}}}^{S}$ are structured similarly, with rows corresponding to either antennas ($A$) or scatterers ($S$), and columns spanning all structures in $A\cup S$. Explicitly, the $pq$-th sub-block of these matrices ($p\in A/S$, $q\in A\cup S$) is given by: 
\begin{equation}
  \hat{\bm{\mathcal{G}}}_{pq}^{A/S}=\left( 1-\delta _{pq} \right) \bm{\mathcal{G}}_{pq}^t
\end{equation}
where the superscript ${}^t$ denotes matrix transpose, and $\delta_{pq}$ is the Kronecker delta ensuring that the diagonal blocks of $\hat{\bm{\mathcal{G}}}^{A/S}$ vanish.

The two latter equations in \eqref{eq11} can be succinctly combined into a single equation:
\begin{equation}
  \label{eq15}
  \begin{bmatrix}
    \hat{\mathbf{T}}^A \hat{\mathbf{v}}^A\\
    \mathbf{0}\\
  \end{bmatrix} + (\hat{\mathbf{S}}-\mathbf{1})\hat{\mathbf{a}}_\mathrm{d} + (\hat{\mathbf{S}}-\mathbf{1}) \hat{\bm{\mathcal{G}}} \hat{\mathbf{f}} = \hat{\mathbf{f}}
\end{equation}
from which the scattered field vector $\hat{\mathbf{f}}$ is directly obtained as
\begin{equation}
  \label{eq16}
  \hat{\mathbf{f}}=\left\{ \mathbf{1}-\left( \hat{\mathbf{S}}-\mathbf{1} \right) \hat{\bm{\mathcal{G}}} \right\} ^{-1}\left\{ \begin{bmatrix}
    \hat{\mathbf{T}}^A \hat{\mathbf{v}}^A\\
    \mathbf{0}\\
  \end{bmatrix} +\left( \hat{\mathbf{S}}-\mathbf{1} \right) \hat{\mathbf{a}}_\mathrm{d} \right\} .
\end{equation}
Here, the vectors and matrices have been structured as:
\begin{equation}
  \hat{\mathbf{a}}_\mathrm{d}=\begin{bmatrix}
    \hat{\mathbf{a}}_\mathrm{d}^{A}\\
    \hat{\mathbf{a}}_\mathrm{d}^{S}
  \end{bmatrix},
  \hat{\bm{\mathcal{G}}}=\begin{bmatrix}
    \hat{\bm{\mathcal{G}}}^A\\
    \hat{\bm{\mathcal{G}}}^S
  \end{bmatrix},
  \hat{\mathbf{S}}=\mathrm{diag}\left( \hat{\mathbf{S}}^A,\hat{\mathbf{S}}^S \right). 
\end{equation}

To further streamline the derivation, we define the auxiliary matrix $\mathbf{M}\equiv\mathbf{1}-\left( \hat{\mathbf{S}}-\mathbf{1} \right) \hat{\bm{\mathcal{G}}}$, partitioning it explicitly into block matrices as
\begin{equation*}
  \mathbf{M}=
  \begin{bmatrix}
    \mathbf{M}_{AA}&		\mathbf{M}_{AS}\\
    \mathbf{M}_{SA}&		\mathbf{M}_{SS}\\
  \end{bmatrix}.
\end{equation*}
Introducing the Schur complement:
\begin{equation}
  \begin{split}
  &\tilde{\mathbf{M}}=\mathbf{M}_{AA}-\mathbf{M}_{AS}\mathbf{M}_{SS}^{-1}\mathbf{M}_{SA}
  \\
  &\mathbf{M}_L=\begin{bmatrix}
    \mathbf{1}&		-\mathbf{M}_{SS}^{-1}\mathbf{M}_{SA}\\
  \end{bmatrix} ^t\tilde{\mathbf{M}}^{-1}
  \end{split}
\end{equation} 
we derive (see Appendix~\ref{app_C} for details)
\begin{equation}
  \label{eq19}
  \mathbf{M} ^{-1} \begin{bmatrix}
    \hat{\mathbf{T}}^A \hat{\mathbf{v}}^A\\
    \mathbf{0}\\
  \end{bmatrix}=\mathbf{M}_L\hat{\mathbf{T}}^A \mathbf{v}^A.
\end{equation}
Hence, \eqref{eq16} can be rewritten as
\begin{equation}
  \label{eq20}
  \hat{\mathbf{f}}=\mathbf{M}_L\hat{\mathbf{T}}^A \hat{\mathbf{v}}^A + \mathbf{M}^{-1}\left( \hat{\mathbf{S}}-\mathbf{1} \right) \hat{\mathbf{a}}_\mathrm{d}.
\end{equation}

Substituting \eqref{eq20} into the first equation of \eqref{eq11}, we arrive at an explicit expression for the guided output wave vector: 
\begin{equation}
  \begin{split}
    \hat{\mathbf{w}}^A&=\left( \hat{\mathbf{\Gamma}}^A+\hat{\mathbf{R}}^A \hat{\bm{\mathcal{G}}} ^A\mathbf{M}_L\hat{\mathbf{T}}^A \right) \hat{\mathbf{v}}^A
    \\
    &\quad +\hat{\mathbf{R}}^A\hat{\mathbf{a}}_\mathrm{d}^{A}+\hat{\mathbf{R}}^A\mathbf{M}^{-1}\left( \hat{\mathbf{S}}-\mathbf{1} \right) \hat{\mathbf{a}}_\mathrm{d}
  \end{split}
\end{equation}
which may also be represented as
\begin{equation}
  \label{eq22}
  \begin{split}
    \hat{\mathbf{w}}^A&=\left( \hat{\mathbf{\Gamma}}^A+\hat{\mathbf{R}}^A \hat{\bm{\mathcal{G}}} ^A\mathbf{M}_L\hat{\mathbf{T}}^A \right) \hat{\mathbf{v}}^A
    \\
    &\quad +\left\{ \left[ \begin{matrix}
      \hat{\mathbf{R}}^A&		\mathbf{0}\\
    \end{matrix} \right] +\hat{\mathbf{R}}^A\mathbf{M}^{-1}\left( \hat{\mathbf{S}}-\mathbf{1} \right) \right\} \hat{\mathbf{a}}_\mathrm{d}.
  \end{split}
\end{equation}

Introducing a unified representation as
\begin{equation}
  \label{eq23}
  \begin{bmatrix}
    \mathbf{\Gamma }_G&		\mathbf{R}_G\\
    \mathbf{T}_G&		\mathbf{S}_G-\mathbf{1}\\
  \end{bmatrix} \begin{bmatrix}
    \hat{\mathbf{v}}^A\\
    \hat{\mathbf{a}}_\mathrm{d}\\
  \end{bmatrix} = \begin{bmatrix}
    \hat{\mathbf{w}}^A\\
    \hat{\mathbf{f}}\\
  \end{bmatrix}
\end{equation}
comparison with equations \eqref{eq20} and \eqref{eq22} allows us to explicitly determine each matrix block as
\begin{equation}
  \label{eq24}
  \begin{split}
    &\mathbf{\Gamma }_G=\hat{\mathbf{\Gamma}}^A+\hat{\mathbf{R}}^A \hat{\bm{\mathcal{G}}}^A\mathbf{M}_L\hat{\mathbf{T}}^A
    \\
    &\mathbf{R}_G=\left[ \begin{matrix}
	    \hat{\mathbf{R}}^A&		\mathbf{0}\\
    \end{matrix} \right] +\hat{\mathbf{R}}^A \hat{\bm{\mathcal{G}}}^A \mathbf{M}^{-1}\left( \hat{\mathbf{S}}-\mathbf{1} \right) 
    \\
    &\mathbf{T}_G=\mathbf{M}_L\hat{\mathbf{T}}^A
    \\
    &\mathbf{S}_G-\mathbf{1}=\mathbf{M}^{-1}\left( \hat{\mathbf{S}}-\mathbf{1} \right).
  \end{split}
\end{equation}
This formulation generalizes previously known cases neatly: For purely antenna-based systems, removing scatterer-related blocks simplifies $\hat{\bm{\mathcal{G}}}=\hat{\bm{\mathcal{G}}}^A$ and $\mathbf{M}_L=\mathbf{M}^{-1}$, thereby recovering results consistent with \cite{ref_3D_FEM,ref_myGSM}. Conversely, for purely scattering systems, only the last equation of \eqref{eq24} persists, corresponding exactly to the multiple-scattering formulation discussed in \cite{ref_Multiple_E3}. However, the fully matrix-based representation developed here is considerably more concise than the series expansions utilized in earlier treatments \cite{ref_Multiple_E3}.

An alternative yet physically insightful representation of \eqref{eq24} can be derived, which, while less computationally efficient, clearly separates coupling effects from the intrinsic responses of individual structures. Replacing $\hat{\mathbf{f}}$ on the left-hand side of \eqref{eq15} with \eqref{eq20}, we obtain
\begin{equation}
  \begin{split}
  \hat{\mathbf{f}}&=\left\{ \begin{bmatrix}
    \hat{\mathbf{T}}^A\\
    \mathbf{0}\\
  \end{bmatrix} +(\hat{\mathbf{S}}-\mathbf{1})\hat{\bm{\mathcal{G}}}\mathbf{M}_L\hat{\mathbf{T}}^A \right\} \hat{\mathbf{v}}^A
  \\
  &\qquad +\left\{ (\hat{\mathbf{S}}-\mathbf{1})+(\hat{\mathbf{S}}-\mathbf{1})\hat{\bm{\mathcal{G}}}\mathbf{M}^{-1}\left( \hat{\mathbf{S}}-\mathbf{1} \right) \right\} \hat{\mathbf{a}}_\mathrm{d}.
  \end{split}
\end{equation}
Consequently, the matrices $\mathbf{T}_G$ and $\mathbf{S}_G-\mathbf{1}$ can be explicitly rewritten as:
\begin{equation}
  \begin{split}
  &\mathbf{T}_G=\begin{bmatrix}
    \hat{\mathbf{T}}^A\\
    \mathbf{0}\\
  \end{bmatrix} +(\hat{\mathbf{S}}-\mathbf{1})\hat{\bm{\mathcal{G}}}\mathbf{M}_L\hat{\mathbf{T}}^A
  \\
  &\mathbf{S}_G-\mathbf{1}=(\hat{\mathbf{S}}-\mathbf{1})+(\hat{\mathbf{S}}-\mathbf{1})\hat{\bm{\mathcal{G}}}\mathbf{M}^{-1}\left( \hat{\mathbf{S}}-\mathbf{1} \right).
  \end{split}
\end{equation}

Combining the above with the first two equations in \eqref{eq24}, we arrive at:
\begin{equation}
  \label{eq27}
  \begin{split}
   &\begin{bmatrix}
    \mathbf{\Gamma }_G&		\mathbf{R}_G\\
    \mathbf{T}_G&		\mathbf{S}_G-\mathbf{1}\\
  \end{bmatrix}  =\begin{bmatrix}
    \hat{\mathbf{\Gamma}}^A&		\hat{\mathbf{R}}^A&		\mathbf{0}\\
    \hat{\mathbf{T}}^A&		\hat{\mathbf{S}}^A-\mathbf{1}&		\mathbf{0}\\
    \mathbf{0}&		\mathbf{0}&		\hat{\mathbf{S}}^S-\mathbf{1}\\
  \end{bmatrix}
  \\
  &\quad +\begin{bmatrix}
    \hat{\mathbf{R}}&		\\
    &		\hat{\mathbf{S}}-\mathbf{1}\\
  \end{bmatrix} \begin{bmatrix}
    \hat{\bm{\mathcal{G}}}^A\\
    \hat{\bm{\mathcal{G}}}\\
  \end{bmatrix} \begin{bmatrix}
    \mathbf{M}_L&		\mathbf{M}^{-1}\\
  \end{bmatrix} \begin{bmatrix}
    \hat{\mathbf{T}}&		\\
    &		\hat{\mathbf{S}}-\mathbf{1}\\
  \end{bmatrix}.
  \end{split}
\end{equation}
In this expression, the first term explicitly represents the isolated intrinsic response of each component, whereas the second term clearly illustrates the coupling interactions among different structures.

However, equations \eqref{eq24} and \eqref{eq27} are not yet directly the GS-matrix of the entire system since vectors $\hat{\mathbf{a}}_\mathrm{d}$ and $\hat{\mathbf{f}}$ are expressed within local coordinate systems of individual components, rather than the global coordinate framework. The total scattered wave from the complete system is the superposition of individual contributions transformed to the global coordinate system, yielding:
\begin{equation}
  \label{eq29}
  \mathbf{f}=\sum_p{\bm{\mathcal{R}}_p \mathbf{f}^p}.
\end{equation}
In compact matrix form, \eqref{eq29} can be expressed as
\begin{equation}
  \label{eq30}
  \mathbf{f}=\hat{\bm{\mathcal{R}}} \hat{\mathbf{f}}
\end{equation}
where the global translation matrix is defined as $\hat{\bm{\mathcal{R}}}=\left[ \bm{\mathcal{R}}_1, \bm{\mathcal{R}}_2,\cdots ,\bm{\mathcal{R}}_{M+N} \right] $. Here, $\bm{\mathcal{R}}_p=\bm{\mathcal{R}}(k\boldsymbol{r}_p)$ represents another translation matrix (see details in Sec.~\ref{Sec_IV}).

Similarly, the direct incident wave vector for the $p$-th structure, $\mathbf{a}_\mathrm{d}^p$, is simply the translation of the global incident wave $\mathbf{a}$ into the local coordinate system, \ie $\mathbf{a}_\mathrm{d}^p=\bm{\mathcal{R}}_p^t\mathbf{a}$. Hence, analogous to \eqref{eq30}, we have
\begin{equation}
  \label{eq31}
  \hat{\mathbf{a}}_\mathrm{d}=\hat{\bm{\mathcal{R}}}^t \mathbf{a}.
\end{equation}

Substituting \eqref{eq30} and \eqref{eq31} into \eqref{eq23} and comparing the result with \eqref{eq3}, the full-system GS-matrix can be deduced (more precisely, its variant):
\begin{equation}
  \label{eq32}
  \begin{bmatrix}
    \mathbf{\Gamma }&		\mathbf{R}\\
    \mathbf{T}&		\mathbf{S}-\mathbf{1}\\
  \end{bmatrix}  =\begin{bmatrix}
    \mathbf{1}&		\\
    &		\hat{\bm{\mathcal{R}}}\\
  \end{bmatrix} \begin{bmatrix}
    \mathbf{\Gamma }_G&		\mathbf{R}_G\\
    \mathbf{T}_G&		\mathbf{S}_G-\mathbf{1}\\
  \end{bmatrix} \begin{bmatrix}
    \mathbf{1}&		\\
    &		\hat{\bm{\mathcal{R}}}^t\\
  \end{bmatrix}.
\end{equation}

In many practical scenarios, explicitly forming the full GS-matrix might not be necessary. Instead, the interest typically lies in evaluating the system's response to given excitations, \eg mapping $\hat{\mathbf{v}}^A$ and $\hat{\mathbf{a}}_\mathrm{d}$ into $\hat{\mathbf{w}}^A$ and $\hat{\mathbf{f}}$, as per \eqref{eq20} and \eqref{eq22}. Due to the rapid convergence of the Neumann series:
\begin{equation}
  \label{eq_Neumann}
  \mathbf{M}^{-1}=\mathbf{1}+\left( \hat{\mathbf{S}}-\mathbf{1} \right) \hat{\bm{\mathcal{G}}}+\left\{ \left( \hat{\mathbf{S}}-1 \right) \hat{\bm{\mathcal{G}}} \right\} ^2+\cdots 
\end{equation}
calculations involving $\mathbf{M}^{-1}$ and $\mathbf{M}_L$ (the left column of the 2$\times$2 matrix-partition form of $\mathbf{M}^{-1}$) in \eqref{eq20} and \eqref{eq22} can efficiently proceed through successive matrix-vector multiplications without explicit matrix inversion. This significantly reduces computational complexity. The peak memory requirement in this computational procedure is mainly dictated by the storage of the dense matrix $\hat{\bm{\mathcal{G}}}$. For large-scale problems, it is beneficial to compute the matrix-vector products directly in a block-wise manner rather than explicitly forming $\hat{\bm{\mathcal{G}}}$. 

It is important to emphasize that this GS-matrix framework is particularly advantageous for analyzing large-scale hybrid systems composed of numerous individually manageable components, rather than arbitrary large-scale problems lacking such a modular structure.

\section{Evaluation of the Translation Matrix}
\label{Sec_IV}

A crucial step in the preceding method is accurately evaluating the translation matrix $\bm{\mathcal{R}}$, which relates the vector spherical wave expansion coefficients between global and local coordinate systems. While general representations suitable for arbitrary translation directions have been provided in \cite{ref_sph_addition1,ref_sph_addition2,ref_scattering_theory,ref_Multiple_E3, ref_Kristensson_booklet}, these formulations can be computationally intensive. A simpler closed-form analytical expression exists only when the translation is along the $z$-axis \cite[Appendix C]{ref_my}. Consequently, a practical approach is to first rotate coordinates to align the translation vector with the new $z$-axis, evaluate the translation matrix efficiently in that direction, and then rotate back to the original coordinates \cite{ref_sph_near_measure,ref_3D_FEM,ref_mysyn_CMA}.

Therefore, the translation matrix $\bm{\mathcal{R}}(k\boldsymbol{d})$ along the vector $\boldsymbol{d}=(d,\theta_d,\varphi_d)$, with $d>0$, $\theta_d\in[0,\pi]$, and $\varphi_d\in[0,2\pi]$, can be represented as
\begin{equation}
  \bm{\mathcal{R}} \left( k\boldsymbol{d} \right) =\bm{\mathcal{D}} _{d}^{t}\bm{\mathcal{R}} ^z\left( kd \right) \bm{\mathcal{D}} _d.
\end{equation}
Here, $\bm{\mathcal{D}}_d = \bm{\mathcal{D}}(\varphi_d, \theta_d, 0)$ is a rotation matrix whose inverse is given by $\bm{\mathcal{D}}^{-1}=\bm{\mathcal{D}}^{t}$. Detailed definitions of $nn'$-th element of $\bm{\mathcal{D}}$ and the z-axis translation matrix $\bm{\mathcal{R}}^z$ can be found in \cite[Appendix B]{ref_mysyn_CMA} and \cite[Appendix C]{ref_my}, respectively. For the reader's convenience, we reproduce these definitions in Appendices \ref{app_A} and \ref{app_B} of this paper. The indices of the VSWFs are denoted compactly by $n\rightarrow \tau\sigma ml$, indicating polarization type $\tau={1,2}$ (TE, TM), parity $\sigma={e,o}$ (even, odd), angular degree $l={1,2,\dots,L_{\max}}$, and order $m={0,1,\dots,l}$.

Another essential translation matrix $\bm{\mathcal{G}}$, which converts the spherical expansion coefficients of scattered waves from one structure into the incident waves at another, can be similarly expressed as:
\begin{equation}
  \bm{\mathcal{G}} \left( k\mathbf{d} \right) =\frac{1}{2}\bm{\mathcal{D}} _{d}^{t}\bm{\mathcal{Y}} ^z\left( kd \right) \bm{\mathcal{D}} _d.
\end{equation}
If the minimal circumscribing spheres around the two structures do not overlap, $\bm{\mathcal{Y}}^{z}$ shares an analytical form identical to that of $\bm{\mathcal{R}}^{z}$, except the spherical Bessel function $j_{\lambda}(kd)$ is replaced by the spherical Hankel function $h_{\lambda}^{(2)}(kd)$.

\begin{figure}[!t]
  \centering
  \includegraphics[]{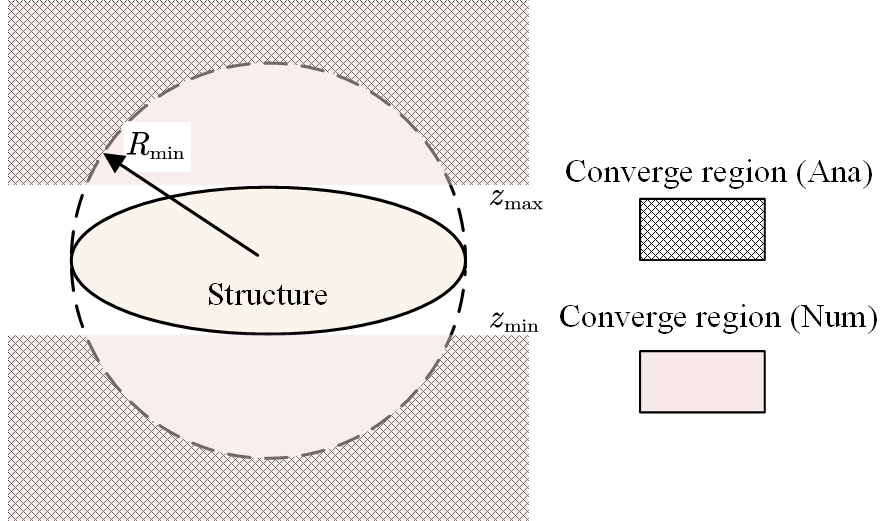}
  \caption{Convergence regions for $z$-axis translations. Analytical representations of $\bm{\mathcal{Y}}^z$ require the other structure to lie in the region $|z|>R_{\min}$, whereas numerical integral representations are valid within a broader region, bounded by $z>z_{\max}$ and $z<z_{\min}$.}
  \label{fConverge_region}
\end{figure}

When structures are placed very closely, integral-form representations of $\bm{\mathcal{Y}}^z$ significantly relax the previous non-overlapping constraint, allowing translations provided that the structures remain separated by at least one intermediate plane \cite{ref_Rubio,ref_translation_PRA}. Fig.~\ref{fConverge_region} illustrates the convergence regions for evaluating $\bm{\mathcal{Y}}^z$ using different methods. Compared to analytical forms, numerical integral representations extend applicability to closely spaced configurations typical of practical scenarios, at the expense of numerical evaluation.

The integral-form expression for the $nn'$-th element of $\bm{\mathcal{Y}}^z$ is given by
\begin{equation}
  \label{eq_Yint}
  \bm{\mathcal{Y}} _{nn^{\prime}}^{z}\left( kd \right) =2\sum_{i=1}^{2}{\int_C{B_{ni}^{\dagger}\left( \hat{\gamma} \right) e^{-\mathrm{j}kd\cos \alpha}B_{n^{\prime}i}\left( \hat{\gamma} \right) \mathrm{d}\hat{\gamma}}}.
\end{equation}
The slight deviation from \cite{ref_Kristensson_booklet} arises from the time convention assumption of $e^{\mathrm{j}\omega t}$. The integral domain is $\int_C{\mathrm{d}\hat{\gamma}}=\int_C{\sin \alpha \mathrm{d}\alpha \int_0^{2\pi}{\mathrm{d}\beta}}$. The functions $ B_{ni}\left( \hat{\gamma} \right)$ are defined as: 
\begin{equation*}
  B_{ni}\left( \hat{\gamma} \right) =B_{ni}\left( u \right) A_{ni}\left( \beta \right) ,\quad u=\cos\alpha
\end{equation*}
with
\begin{subequations} \label{eq_Bnj}
  \begin{align}
    &B_{ni}\left( u \right) =-\mathrm{j}^{-l}\left[ \delta _{\tau i}\mathrm{j}\Delta _{l}^{m}\left( u \right) +\delta _{\bar\tau i}\pi _{l}^{m}\left( u \right) \right] 
    \label{eq_Bnj_a}
    \\
    &A_{ni}\left( \beta \right) =\delta _{\tau i}\begin{cases}
      \cos \left( m\beta \right)\\
      \sin \left( m\beta \right)\\
    \end{cases}+\delta _{\bar\tau i}\begin{cases}
      -\sin \left( m\beta \right)\\
      \cos \left( m\beta \right)\\
    \end{cases}.
  \end{align}
\end{subequations}
Here, the upper (lower) form of $A_{ni}(\beta)$ applies for even (odd) parity $\sigma=e(o)$. $\bar\tau=3-\tau$. The auxiliary functions
\begin{equation}
  \begin{split}
    &\Delta _{l}^{m}\left( u \right) =-\frac{\sqrt{1-u^2}}{\sqrt{l\left( l+1 \right)}}\frac{\mathrm{d}}{\mathrm{d}u}\tilde{P}_{l}^{m}\left( u \right) \\
    &\pi _{l}^{m}\left( u \right) =-\frac{m}{\sqrt{l\left( l+1 \right)}\sqrt{1-u^2}}\tilde{P}_{l}^{m}\left( u \right)
  \end{split}
\end{equation}
where $\tilde{P}_l^m(u)$ denotes the normalized associated Legendre functions. The superscript $\dagger$ on $B_{ni}^{\dagger}(u)$ denotes conjugation by replacing every explicit $\mathrm{j}$ with $-\mathrm{j}$ in \eqref{eq_Bnj_a}.

The integration over the angle $\beta$ in \eqref{eq_Yint} admits closed-form expressions
\begin{equation*}
  \int_0^{2\pi}{A_{ni}\left( \beta \right) A_{n^{\prime}i}\left( \beta \right)}\mathrm{d}\beta =\delta _{mm^\prime}I_{\tau \sigma ,\tau ^{\prime}\sigma ^{\prime}}^{i}
\end{equation*}
where $I_{\tau\sigma,\tau'\sigma'}^{i}$ takes nonzero values only for the following specific conditions: 
\begin{equation*}
  \begin{split}
    &I_{\tau \sigma ,\tau \sigma}^{i}=\pi \left[ 1+\left( -1 \right) ^{i+\tau +\sigma}\delta _{m0} \right]\\
    &I_{\tau \sigma ,\tau ^{\prime}\sigma ^{\prime}}^{i}=\pi \left[ \left( -1 \right) ^{i+\tau +\sigma}+\delta _{m0} \right] ,\tau \ne \tau ^\prime\land \sigma \ne \sigma ^\prime
  \end{split}
\end{equation*}

Thus, \eqref{eq_Yint} simplifies significantly to a single integral expression (setting $u=\cos\alpha$):
\begin{equation}
  \label{eq_YInt_Final}
  \begin{split}
    \bm{\mathcal{Y}} _{nn^{\prime}}^{z}\left( kd \right) &=2\delta _{mm^\prime}\times\\
    &\sum_{i}^{\left\{ 1,2 \right\}}{I_{\tau \sigma ,\tau ^{\prime}\sigma ^{\prime}}^{i}\int_{u_m}^1{B_{ni}^{\dagger}\left( u \right) e^{-\mathrm{j}kdu}B_{n^{\prime}i}\left( u \right) \mathrm{d}u}}.
  \end{split}
\end{equation}
The theoretical lower bound of \eqref{eq_YInt_Final} should be taken as $u_m=-\mathrm{j}\infty$, leading to convergence to its analytical form \cite{ref_Kristensson_booklet}. However, to accommodate wave translations between closely spaced structures, the lower limit is truncated to a very finite value $u_m = \sqrt{1 - \tilde{\kappa}_m^2}$ (with the convention throughout this paper that $\sqrt{-1}\equiv-\mathrm{j}$). Here, $\tilde{\kappa}_m$ is a small real number greater than one \cite{ref_translation_PRA}. Within the range $0.5 \le kR_{\min} \le 10, L_{\max} \le 20$, $\tilde{\kappa}_m$ may be empirically approximated as \cite{ref_PRA_truncation_number}:
\begin{equation}
  \tilde{\kappa}_m=\left( 0.38L_{\max}+1 \right) \left( kR_{\min} \right) ^{-1}+0.03kR_{\min}.
\end{equation}

The integral in \eqref{eq_YInt_Final} can be efficiently evaluated numerically using a one-dimensional Gauss-Legendre quadrature. Several useful symmetry relations further reduce computational demands: 
\begin{equation}
  \begin{split}
    &\bm{\mathcal{Y}} _{nn^{\prime}}\left( -k\boldsymbol{d} \right) =\bm{\mathcal{Y}} _{n^{\prime}n}\left( k\boldsymbol{d} \right) \\
    &\bm{\mathcal{Y}} _{n^{\prime}n}\left( k\boldsymbol{d} \right) =\left( -1 \right) ^{\tau +\tau ^{\prime}+l-l^{\prime}}\bm{\mathcal{Y}} _{nn^{\prime}}\left( k\boldsymbol{d} \right) 
  \end{split}
\end{equation}
valid for both analytical and integral representations. These identities also apply directly to translation matrices $\bm{\mathcal{G}}$ and $\bm{\mathcal{R}}$.

\section{Numerical Examples}
\subsection{Example 1---Demonstration of Convergence Region for the Translation Matrix}

\begin{figure}[!t]
  \centering
  \subfloat[]{
    \begin{tikzpicture}
      \node (img1) {\includegraphics{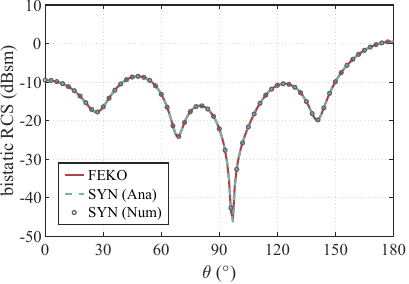}};
      \node at (img1.center)[shift={(2.1cm,-0.3cm)}]{\includegraphics{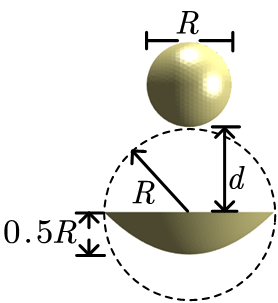}};
      \label{fRCS_alpha}
    \end{tikzpicture} 
  }
  \vfil
  \subfloat[]{\includegraphics[]{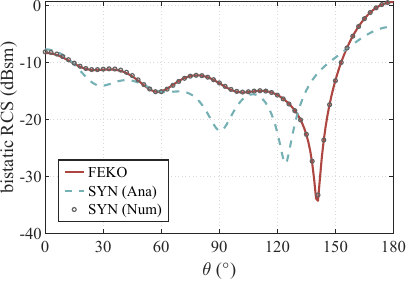}}
  \vfil
  \subfloat[]{\includegraphics[]{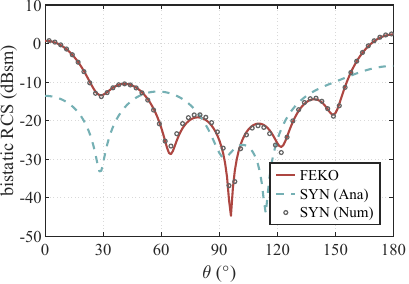}}
  \caption{Bistatic RCS of a PEC paraboloid-sphere system at $kR=2\pi$. (a) At $d=R$, both synthesized methods (`SYN (Ana)' and `SYN (Num)') accurately match FEKO results. (b) At $d=0.1R$, the analytical translation (`SYN (Ana)') diverges, while the numerical integration (`SYN (Num)') remains convergent. (c) At the critical spacing $d=0$, the numerical method (`SYN (Num)') begins to show deviations. Computations were performed with $L_{\max}=20,\tilde{\kappa}_m=1.8$.}
\label{f_RCS_paraboloid}
\end{figure}

To illustrate clearly the convergence characteristics of different translation matrix evaluation methods, we consider the bistatic radar cross section (RCS) of a dual-structure system composed of a perfect electric conductor (PEC) sphere and a PEC paraboloid. The initial configuration is depicted in Fig.~\ref{f_RCS_paraboloid}a, where the PEC sphere is positioned immediately outside the minimal circumscribing sphere of the paraboloid. For this arrangement, the GS-matrix of the combined structure was synthesized according to the theory developed in Sec. \ref{Sec_III}. The computed bistatic RCS under a top-directed plane-wave illumination is shown alongside reference results obtained via FEKO full-wave simulation. The synthesized results employing both analytical (`SYN (Ana)') and numerical integral (`SYN (Num)') translation matrices match perfectly with the FEKO simulation, demonstrating successful convergence in both cases.

Next, we reduce the separation distance $d$ between the sphere and the paraboloid to $0.1R$. The corresponding bistatic RCS results are presented in Fig.~\ref{f_RCS_paraboloid}b. Since the PEC sphere now overlaps with the minimal circumscribing sphere of the paraboloid, the analytical translation matrix-based results (`SYN (Ana)') diverge significantly, whereas the numerical integral-based method (`SYN (Num)') remains accurate and matches FEKO results closely, confirming its broader convergence region.

The critical limiting case occurs at $d=0$, where the sphere's bottom just contacts the paraboloid's aperture. In this extreme scenario, shown in Fig.~\ref{f_RCS_paraboloid}c, the numerical integral method (`SYN (Num)') begins to display noticeable deviations, yet it still closely follows the FEKO results. This behavior clearly supports the convergence regions summarized earlier in Fig.~\ref{fConverge_region}.

\subsection{Example 2---Pure scattering system composed of spheres}

\begin{figure}[!t]
  \centering
  \includegraphics[]{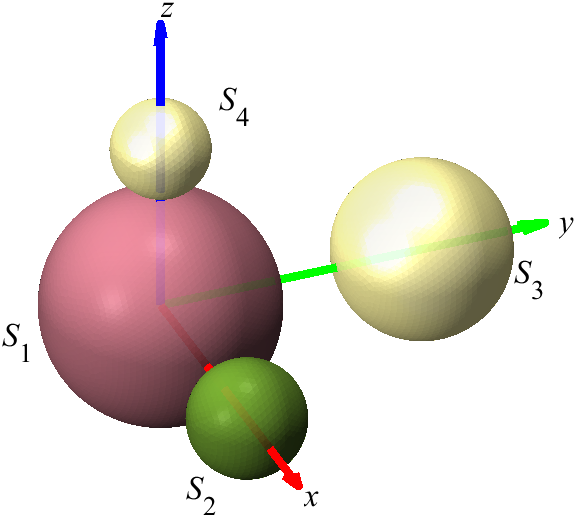}
  \caption{Configuration of a pure scattering system comprising four spheres with varying dimensions and materials.}
  \label{fFour_Sphsca_Model}
\end{figure}

A notable special case arises when all scattering structures are spherical, the scattering matrices are diagonal and can be analytically derived \cite{ref_multiple_spheres,ref_Multiple1,ref_Multiple2,ref_Multiple3,ref_Multiple_E3}. We examine here a scattering scenario composed of four spheres with varying sizes and material properties, as illustrated in Fig.~\ref{fFour_Sphsca_Model}. The spheres $S_1$ to $S_4$ have radii of 24 mm, 12 mm, 18 mm, and 10 mm, positioned at the origin, along the x-axis, y-axis, and z-axis, respectively. Spheres $S_3$ and $S_4$ are PEC, while spheres $S_1$ and $S_2$ are dielectric materials with relative permittivities of $8$ and $4.4 - 8.8\mathrm{j}$, respectively.

\begin{figure}[!t]
  \centering
  \includegraphics[]{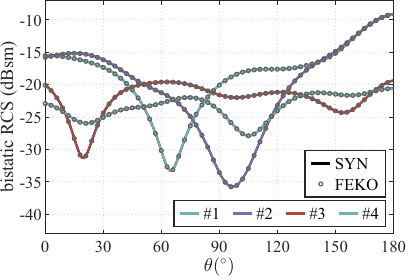}
  \caption{Bistatic RCS of the four-sphere scattering system from Fig.~\ref{fFour_Sphsca_Model}. Curves represent results on the $xoz$-plane. Illumination conditions: (\#1) plane wave incident from $-z$ direction with $x$ polarization, (\#2) $-z$ direction with $y$ polarization, (\#3) $-x$ direction with $z$ polarization, and (\#4) $-y$ direction with $z$ polarization.}
  \label{fFour_Sphsca_RCS}
\end{figure}

Figure~\ref{fFour_Sphsca_RCS} depicts the computed bistatic RCS of this system under plane-wave illumination from several directions and polarizations. The synthesized results using the method developed in Sec.~\ref{Sec_III} match perfectly with full-wave simulations performed using FEKO, validating the effectiveness and accuracy of the proposed synthesis method for multi-sphere scattering problems.

\subsection{Example 3---Pure antenna system}

\begin{figure}[!t]
  \centering
  \includegraphics[]{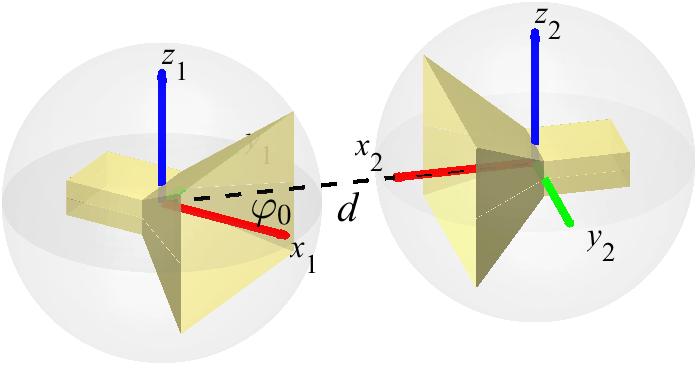}
  \caption{Two-antenna system comprising identical pyramidal horn antennas. Each antenna consists of a rectangular waveguide section (cuboid) and a flared horn aperture. Dimensions of each horn: front aperture 20~cm $\times$ 14.3~cm, rear aperture 7.1~cm $\times$ 3.5~cm, flare length $8,\mathrm{cm}$, and waveguide length 8~cm with cross-section 7.1~cm $\times$ 3.5~cm.}
  \label{f_Horn_dual_model}
\end{figure}

We now consider a pure antenna system consisting of two pyramidal horn antennas, as shown in Fig.~\ref{f_Horn_dual_model}. Both antennas have identical dimensions, separated by a distance $d=45$~cm. The angle between their principal axes (axes $x_1$ and $x_2$) is defined as $\alpha_0=\pi-\varphi_0$.

Following the procedure described in Sec.~\ref{Sec_III}, the synthesis of the overall system's GS-matrix first requires determining the GS-matrix for each individual horn antenna. Because both antennas are identical, their individual GS-matrices are also identical when defined in their respective local coordinate systems. However, the second antenna's coordinate system is rotated by an angle $\alpha_0$ about the $z$-axis relative to the first antenna's coordinate system. To ensure consistency with the theory in Sec.~\ref{Sec_III}, we introduce a coordinate rotation to align the second antenna's local axes with those of the first antenna's coordinate system $(x_1,y_1,z_1)$.

The vectors representing spherical wave expansions before (unprimed) and after (primed) rotation are related by
\begin{equation}
  \mathbf{f}'=\bm{\mathcal{D}}_r^t \mathbf{f},\mathbf{a}'=\bm{\mathcal{D}}_r^t \mathbf{a} \text{ or }
  \mathbf{f}=\bm{\mathcal{D}}_r \mathbf{f}',\mathbf{a}=\bm{\mathcal{D}}_r \mathbf{a}'
\end{equation}
where the spherical-wave rotation matrix $\bm{\mathcal{D}}_r=\bm{\mathcal{D}}(\alpha_0,0,0)$. Consequently, the GS-matrix of the second antenna after rotation can be conveniently expressed as:
\begin{equation}
  \label{eq34}
  \tilde{\mathbf{S}}^{\prime}=\begin{bmatrix}
    \mathbf{1}&		\\
    &		\bm{\mathcal{D}}_r^t\\
  \end{bmatrix}  \tilde{\mathbf{S}} \begin{bmatrix}
    \mathbf{1}&		\\
    &		\bm{\mathcal{D}}_r\\
  \end{bmatrix}.
\end{equation}
This relation arises naturally because
\begin{equation*}
  \begin{split}
  \begin{bmatrix}
    \mathbf{w}^{\prime}\\
    \mathbf{f}^{\prime}\\
  \end{bmatrix} &= \begin{bmatrix}
    \mathbf{1}&		\\
    &		\bm{\mathcal{D}}_r^t\\
  \end{bmatrix}  \begin{bmatrix}
    \mathbf{w}\\
    \mathbf{f}\\
  \end{bmatrix}
  \\
  &= \begin{bmatrix}
    \mathbf{1}&		\\
    &		\bm{\mathcal{D}}_r^t\\
  \end{bmatrix} \tilde{\mathbf{S}} \begin{bmatrix}
    \mathbf{v}\\
    \mathbf{a}\\
  \end{bmatrix} = \begin{bmatrix}
    \mathbf{1}&		\\
    &		\bm{\mathcal{D}}_r^t\\
  \end{bmatrix} \tilde{\mathbf{S}} \begin{bmatrix}
    \mathbf{1}&		\\
    &		\bm{\mathcal{D}}_r\\
  \end{bmatrix} \begin{bmatrix}
    \mathbf{v}^{\prime}\\
    \mathbf{a}^{\prime}\\
  \end{bmatrix}.
  \end{split}
\end{equation*}

Equation \eqref{eq34} offers substantial computational efficiency compared to recomputing or simulating the second antenna, as the computational overhead of determining the rotation matrix is minimal. This efficiency significantly facilitates parametric studies involving antenna orientation adjustments.

\begin{figure}[!t]
  \centering
  \subfloat[]{\includegraphics[]{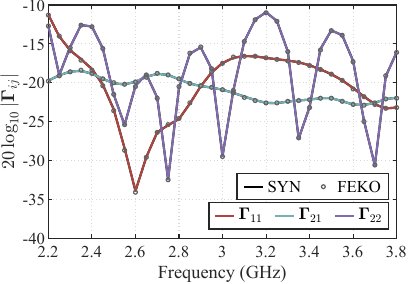}}
  \vfil
  \subfloat[]{\includegraphics[]{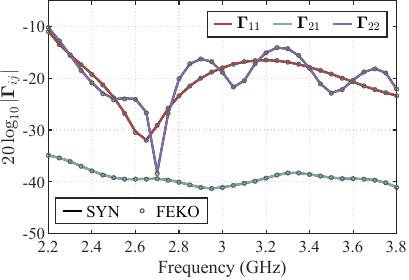}}
  \vfil
  \subfloat[]{\includegraphics[]{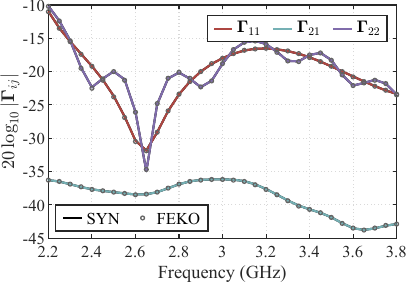}}
  \caption{S-parameters for two pyramidal horn antennas at varying orientation angles $\varphi_0$: (a) $\varphi_0=45^\circ$, (b) $\varphi_0=90^\circ$, and (c) $\varphi_0=135^\circ$.}
\label{f_Spara_DualHorn_vary_phi}
\end{figure}

After obtaining the rotated GS-matrix of the second antenna through \eqref{eq34}, the overall GS-matrix for the two-antenna system was synthesized following the procedure detailed in Sec.~\ref{Sec_III}. Fig.~\ref{f_Spara_DualHorn_vary_phi} presents the computed S-parameters as the angular $\varphi_0$ between the two antennas varies. The resulting variation aligns well with engineering intuition: as $\varphi_0$ increases, antenna 2 increasingly moves away from the main lobe of antenna 1, thus decreasing its received power. Furthermore, rotating antenna 2 from $\varphi_0=90^\circ$ to $\varphi_0=135^\circ$ minimally affects the reflection coefficient at antenna 1, indicating that antenna 2 no longer significantly obstructs antenna 1. The synthesized results closely agree with FEKO full-wave simulations, verifying the effectiveness and accuracy of the proposed GS-matrix synthesis method for pure antenna systems.

In the computational example above (at $\varphi_0=45^\circ$ and frequency $f=3.8$ GHz), each pyramidal horn antenna was discretized into 5766 RWG basis functions, and 646 VSWFs were employed to construct its GS-matrix. A direct full-wave simulation in FEKO, involving both antennas simultaneously, required a total runtime of approximately 189~s. In contrast, the proposed method, which only required simulation of a single horn antenna, completed in 67~s (comprising 63~s for GS-matrix generation and 4~s for computing mutual couplings). Although a single calculation might not fully reflect the computational advantage of our method, significant efficiency emerges when studying antenna placement effects, such as varying orientation angle $\varphi_0$ at the same frequency. In such scenarios, the GS-matrix is calculated only once; subsequent computations for each new orientation require merely about 4~s, whereas FEKO would need to perform a full recalculation for each orientation.

\begin{figure}[!t]
  \centering
  \subfloat[]{\includegraphics[]{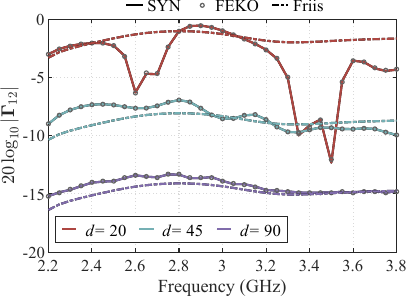}} 
  \vfil
  \subfloat[]{\includegraphics[]{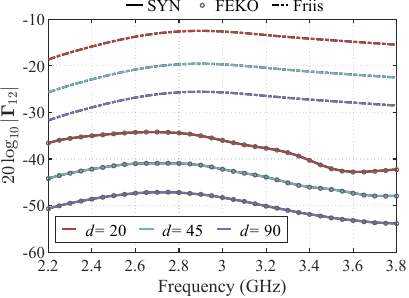}}
  \caption{Transmission coefficients for varying antenna separations $d$ (cm): (a) horn-to-horn antenna transmission; (b) horn-to-dipole antenna transmission. Results from the synthesized GS-matrix (SYN), FEKO simulation, and Friis' formula are compared.} 
\label{f_Spara_Horn_Dipole_vary_d}
\end{figure}

This computational advantage is similarly evident when varying the distance $d$ between antennas. Fig.~\ref{f_Spara_Horn_Dipole_vary_d}a presents the transmission coefficient between two identical pyramidal horn antennas facing each other ($\varphi_0=0$) at different separation distances. The synthesized GS-matrix-based results match closely with FEKO full-wave simulation data. Friis' transmission formula, included for comparison, yields accurate results only at relatively large antenna separations (\eg $d=90$ cm, $f>3.4$ GHz). However, its accuracy deteriorates significantly at smaller distances, likely because Friis' formula is fundamentally a far-field approximation.

Moreover, our method does not necessitate that the antennas be identical or similar. For example, replacing one horn antenna with a dipole still results in excellent agreement between the synthesized results and FEKO full-wave simulations, as demonstrated in Fig.~\ref{f_Spara_Horn_Dipole_vary_d}b. In this case, the dipole antenna, oriented orthogonally to the polarization of the horn, leads to minimal power transmission. By contrast, Friis' formula, which lacks polarization information, significantly overestimates the transmission coefficient.

\subsection{Example 4---A general antenna-scatterer hybrid system}

\begin{figure}[!t]
  \centering
  \includegraphics[]{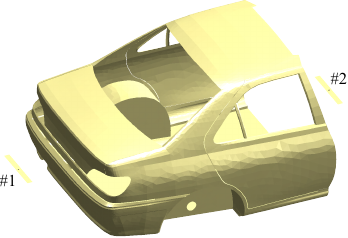}
  \caption{Hybrid antenna-scatterer system comprising two 300~MHz dipole antennas (\#1 and \#2) positioned at the front and rear sides of a car back model. Antenna \#1 is located $0.346\lambda$ from the rear of the car, and the two dipoles are separated by $2.8\lambda$.}
  \label{f_car_2dipole_model}
\end{figure}

The final example demonstrates the validity of our proposed formulation for a realistic hybrid antenna-scatterer scenario. Fig.~\ref{f_car_2dipole_model} illustrates the considered setup, comprising two dipole antennas operating at 300~MHz, positioned along the longitudinal axis with a detailed vehicle model representing the car's rear portion situated between them. The vehicle and each dipole antenna are discretized using 14,754 and 1,400 RWG basis functions, respectively. For the spherical wave expansion, 646 VSWFs are employed. The corresponding FEKO simulation file is publicly accessible via~\cite{ref_github}.

We examine the influence of the vehicle orientation angle ($\alpha$) on the S-parameters of the dipole antennas. Table~\ref{T_spara_car_2dipole} compares synthesized results derived from our GS-matrix-based theory (Sec.~\ref{Sec_III}) with those from FEKO simulations. The comparison reveals excellent agreement, with maximum deviations below 0.3 dB.

\begin{figure}[!t]
  \centering
  \subfloat[]{\includegraphics[]{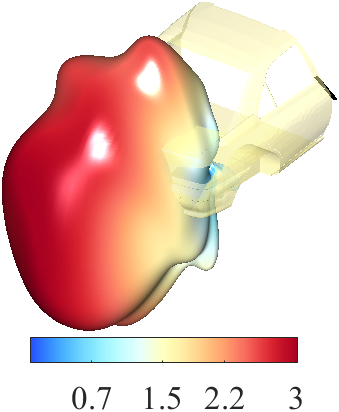}}
  \hfil
  \subfloat[]{\includegraphics[]{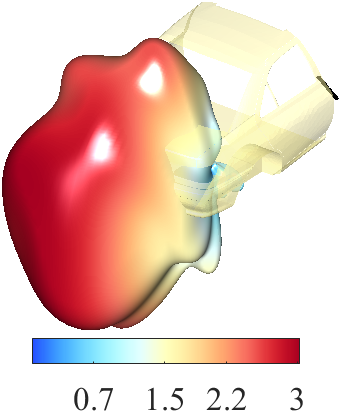}}
  \vfil
  \subfloat[]{\includegraphics[]{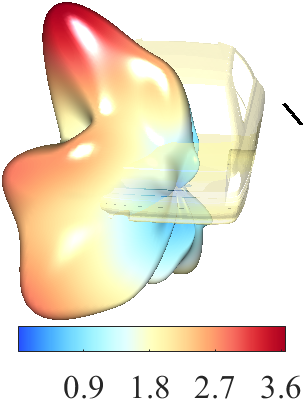}}
  \hfil
  \subfloat[]{\includegraphics[]{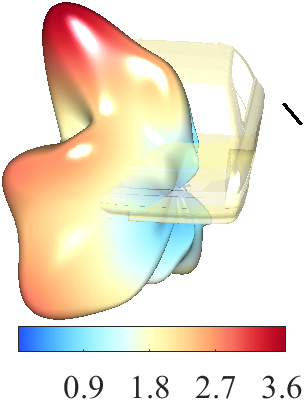}}
  \caption{Gain patterns for dipole antenna \#1 excited alone, with car rotated at (a), (b) $\alpha=30^\circ$ and (c), (d) $\alpha=60^\circ$. Left panels are synthesized GS-matrix results, right panels are FEKO simulations.} 
\label{f_Gain_car_2dipole}
\end{figure}

Further validation is provided by examining antenna radiation patterns when only the first dipole antenna is excited. As depicted in Fig.~\ref{f_Gain_car_2dipole}, the vehicle's scattering significantly distorts the antenna radiation patterns, and this distortion varies notably with the vehicle orientation angle. The synthesized GS-matrix-based patterns closely match those obtained from FEKO simulations. These results underscore the accuracy and practical effectiveness of our method, including the correct implementation of translations and rotations for hybrid systems.

\begin{table}[]
  \centering
  \caption{S-parameters between two dipoles vs. varying car directions}
  \begin{tabular}{ccccccc}
  \midrule
  \multirow{2}{*}{$\alpha (^\circ)$} & \multicolumn{3}{c}{SYN (dB)}  & \multicolumn{3}{c}{FEKO (dB)} \\ \cline{2-7}
                                    &\rule{0pt}{2.5ex} $\mathbf{\Gamma}_{11}$    & $\mathbf{\Gamma}_{22}$    & $\mathbf{\Gamma}_{12}$    & $\mathbf{\Gamma}_{11}$    & $\mathbf{\Gamma}_{22}$    & $\mathbf{\Gamma}_{12}$   \\   \midrule

  0                                 & -10.83 & -13.32 & -42.48 & -10.82 & -13.29 & -42.46 \\
  30                                & -13.11 & -12.95 & -42.78 & -13.09 & -12.86 & -42.88 \\
  60                                & -11.39 & -11.08 & -51.16 & -11.34 & -11.04 & -51.01 \\
  90                                & -16.04 & -16.05 & -43.58 & -15.83 & -15.83 & -43.56 \\
  120                               & -11.10 & -11.37 & -51.65 & -11.03 & -11.32 & -51.47 \\
  150                               & -12.63 & -13.01 & -40.88 & -12.54 & -12.99 & -41.10 \\
  180                               & -13.32 & -10.93 & -42.68 & -13.29 & -10.82 & -42.46 \\
  210                               & -12.82 & -13.10 & -42.90 & -12.79 & -13.08 & -43.01 \\
  240                               & -11.11 & -11.42 & -51.38 & -11.03 & -11.36 & -51.66 \\
  270                               & -16.02 & -16.01 & -43.66 & -15.80 & -15.79 & -43.70 \\
  300                               & -11.38 & -11.06 & -52.50 & -11.34 & -11.02 & -52.13 \\
  330                               & -13.01 & -12.47 & -41.13 & -12.98 & -12.46 & -41.27 \\
  \midrule
  \end{tabular}
  \label{T_spara_car_2dipole}
\end{table}

\section{Conclusion}

This paper has presented a GS-matrix synthesis framework for hybrid electromagnetic systems comprising arbitrary numbers of antennas and scatterers. The proposed method assembles the overall system response from independently obtained S- and GS-matrices of individual substructures, enabling a modular and unified region decomposition approach. It seamlessly accommodates pure antenna systems, pure scatterer systems, and their hybrids, without imposing restrictions on the number or type of components. Although the method requires that constituent structures be separable by at least one plane, this condition is sufficiently general to encompass a broad class of practical multi-structure configurations.

Extensive numerical examples have verified the accuracy, efficiency, and versatility of the formulation across various scenarios. Since the GS-matrix synthesis is agnostic to the underlying numerical solver, the approach can be readily integrated with existing simulation tools, providing a flexible frequency-domain framework for analyzing complex electromagnetic assemblies while significantly reducing computational overhead.

Furthermore, the method naturally supports structural reconfiguration and reuse. In particular, by leveraging the rotation theorem of vector spherical wavefunctions, it enables efficient evaluation of different antenna orientations and polarization settings without requiring repeated full-wave simulations. This makes the proposed technique especially valuable for parametric studies and layout optimization involving modular antenna-scatterer systems.

\begin{appendices}

\section{The Inverse of Matrix in Block-partition Form}
\label{app_C}
For the matrix used in Sec.~\ref{Sec_III} defined as
\begin{equation*}
  \mathbf{M}=\mathbf{1}-\left( \hat{\mathbf{S}}-\mathbf{1} \right) \hat{\bm{\mathcal{G}}}=
  \begin{bmatrix}
    \mathbf{M}_{AA}&		\mathbf{M}_{AS}\\
    \mathbf{M}_{SA}&		\mathbf{M}_{SS}\\
  \end{bmatrix}
\end{equation*}
its inverse, $\mathbf{M}^{-1}$, expressed in block-partitioned form, is:
\begin{equation}
  \label{eq_b3}
  \begin{bmatrix}
    \tilde{\mathbf{M}}^{-1}&		-\tilde{\mathbf{M}}^{-1}\mathbf{M}_{AS}\mathbf{M}_{SS}^{-1}\\
    -\mathbf{M}_{SS}^{-1}\mathbf{M}_{SA}\tilde{\mathbf{M}}^{-1}&		\mathbf{M}_{SS}^{-1}+\mathbf{M}_{SS}^{-1}\mathbf{M}_{SA}\tilde{\mathbf{M}}^{-1}\mathbf{M}_{AS}\mathbf{M}_{SS}^{-1}\\
  \end{bmatrix}
\end{equation}
where $\tilde{\mathbf{M}}=\mathbf{M}_{AA}-\mathbf{M}_{AS}\mathbf{M}_{SS}^{-1}\mathbf{M}_{SA}$ is the Schur complement of $\mathbf{M}_{SS}$ in $\mathbf{M}$. All matrix inversions involved in \eqref{eq_b3} are performed on smaller submatrices compared to the original $\mathbf{M}$.

The left column of $\mathbf{M}^{-1}$ reads
\begin{equation}
  \mathbf{M}_L = \begin{bmatrix}
    \mathbf{1}\\
    -\mathbf{M}_{SS}^{-1}\mathbf{M}_{SA}\\
  \end{bmatrix} \tilde{\mathbf{M}}^{-1}
\end{equation}
which directly yields equation \eqref{eq19}:
\begin{equation*}
  \mathbf{M} ^{-1} \begin{bmatrix}
    \hat{\mathbf{T}}^A \hat{\mathbf{v}}^A\\
    \mathbf{0}\\
  \end{bmatrix}=\mathbf{M}_L\hat{\mathbf{T}}^A \hat{\mathbf{v}}^A.
\end{equation*}

\section{Rotation Matrix}
\label{app_A}

When employing real-valued spherical harmonics to form VSWFs, the rotation matrix $\bm{\mathcal{D}}(\alpha,\beta,\gamma)$ is real-valued and defined by Euler angles $\alpha,\beta,\gamma$. Its $nn'$-th element (with indices $n\rightarrow\tau\sigma ml$) is given by \cite[Appendix B]{ref_mysyn_CMA}: 
\begin{equation}
  \label{eq44}
  \begin{split}
    \bm{\mathcal{D}} _{nn'} =\delta _{\tau \tau '}&\delta _{ll'}\sqrt{\frac{\varepsilon _m\varepsilon _{m'}}{4}}\left( -1 \right) ^{m+m'}
\\
&\times \left[ \begin{matrix}
	\cos \left( m\gamma \right)&		\sin \left( m\gamma \right)\\
	-\sin \left( m\gamma \right)&		\cos \left( m\gamma \right)\\
\end{matrix} \right] 
\\
&\times \left[ \begin{matrix}
	A_{mm'}^{l}\left( \beta \right)&		\\
	&		B_{mm'}^{l}\left( \beta \right)\\
\end{matrix} \right] 
\\
&\times \left[ \begin{matrix}
	\cos \left( m'\alpha \right)&		\sin \left( m'\alpha \right)\\
	-\sin \left( m'\alpha \right)&		\cos \left( m'\alpha \right)\\
\end{matrix} \right] .
\end{split}
\end{equation}
The product in \eqref{eq44} results in a $2\times2$ matrix indexed by $\sigma\sigma'$, where $\sigma,\sigma'={e,o}$. Specifically, $(ee, eo, oe, oo)$ correspond respectively to the $(11, 12, 21, 22)$ elements. The coefficients $A_{mm'}^{l}(\beta)$ and $B_{mm'}^{l}(\beta)$ are: 
\begin{equation}
  \begin{split}
    &A_{mm'}^{l}\left( \beta \right) =d_{mm'}^{l}\left( \beta \right) +\left( -1 \right) ^{m'}d_{m-m'}^{l}\left( \beta \right) \\
    &B_{mm'}^{l}\left( \beta \right) =d_{mm'}^{l}\left( \beta \right) -\left( -1 \right) ^{m'}d_{m-m'}^{l}\left( \beta \right) 
  \end{split}
\end{equation}
where
\begin{equation}
  \begin{split}
    d_{mm'}^{l}\left( \beta \right) &=\sqrt{\frac{\left( l+m \right) !\left( l-m \right) !}{\left( l+m' \right) !\left( l-m' \right) !}}\\
    &\times \cos ^{m+m'}\left( \frac{\beta}{2} \right) \sin ^{m-m'}\left( \frac{\beta}{2} \right)\\
    &\times P_{l-m}^{\left( m-m',m+m' \right)}\left( \cos \beta \right).
  \end{split}
\end{equation}
Here, $P_{n}^{\left( a,b \right)}(x)$ denotes the Jacobi polynomials.

\section{Translation Matrix}
\label{app_B}
The $nn'$-th element of the z-axis translation matrix $\bm{\mathcal{R}}^z(kd)$ is also real and expressed as \cite[Appendix C]{ref_my}:
\begin{equation}
  \label{eq_a2}
  \begin{array}{l}
    \bm{\mathcal{R}}^z _{1\sigma ml,1\sigma ml^{\prime}}=\left[ \left( -1 \right) ^m+\delta _{m0}\left( -1 \right) ^{\sigma} \right]\cdot C_{l,l^{\prime},m}(kd)\\
    \bm{\mathcal{R}}^z _{1\sigma ml,2\sigma ^{\prime}ml^{\prime}}=(-1)^{\sigma+m}D_{l,l^{\prime},m}(kd),\sigma\ne\sigma^\prime\\
    \bm{\mathcal{R}}^z _{2\sigma ml,\tau \sigma ^{\prime}ml^{\prime}}=\bm{\mathcal{R}}^z _{1\sigma ml,\bar\tau\sigma ^{\prime}ml^{\prime}},\tau =1,2\\
    \bm{\mathcal{R}}^z _{nn^\prime}=0,\text{others}.
  \end{array}
\end{equation}
The coefficients $C_{l,l^\prime,m}$, $D_{l,l^\prime,m}$ are explicitly given by:
\begin{equation}
  \begin{aligned}
    C_{l,l^{\prime},m}(kd)=&\frac{\varepsilon_m}{4}\times \sum_{\lambda =\left| l-l^{\prime} \right|}^{l+l^{\prime}}(-1)^{\left( l^{\prime}-l+\lambda \right) /2}(2\lambda +1)\\
    &\times \sqrt{\frac{(2l+1)\left( 2l^{\prime}+1 \right)}{l(l+1)l^{\prime}\left( l^{\prime}+1 \right)}}\\
    &\times \left( \begin{matrix}
    l&		l^{\prime}&		\lambda\\
    0&		0&		0\\
  \end{matrix} \right) \left( \begin{matrix}
    l&		l^{\prime}&		\lambda\\
    m&		-m&		0\\
  \end{matrix} \right)\\
    &\times \left[ l(l+1)+l^{\prime}\left( l^{\prime}+1 \right) -\lambda (\lambda +1) \right]\\
    &\times j_{\lambda}(kd)\\
  \end{aligned}
\end{equation}
and 
\begin{equation}
  \begin{aligned}
    D_{l,l^{\prime},m}(kd)&=-m\times \sum_{\lambda =\left| l-l^{\prime} \right|}^{l+l^{\prime}}(-1)^{\left( l^{\prime}-l+\lambda \right) /2}(2\lambda +1)\\
    &\times \sqrt{\frac{(2l+1)\left( 2l^{\prime}+1 \right)}{l(l+1)l^{\prime}\left( l^{\prime}+1 \right)}}\\
    &\times \left( \begin{matrix}
    l&		l^{\prime}&		\lambda\\
    0&		0&		0\\
  \end{matrix} \right) \left( \begin{matrix}
    l&		l^{\prime}&		\lambda\\
    m&		-m&		0\\
  \end{matrix} \right)\\   
    &\times kd\times j_{\lambda}(kd)\\
  \end{aligned}
\end{equation}
where $\varepsilon_m=2-\delta_{m0}$, $j_\lambda$ denotes the spherical Bessel function, and $$\left( \begin{matrix}
	\cdot&		\cdot&		\cdot\\
	\cdot&		\cdot&		\cdot\\
\end{matrix} \right)$$ represents the Wigner 3-j symbol \cite{ref_Wigner}, computed efficiently via three-term recurrence relations.

\end{appendices}

  \begin{IEEEbiography}[{\includegraphics[width=1in,height=1.25in,clip,keepaspectratio]{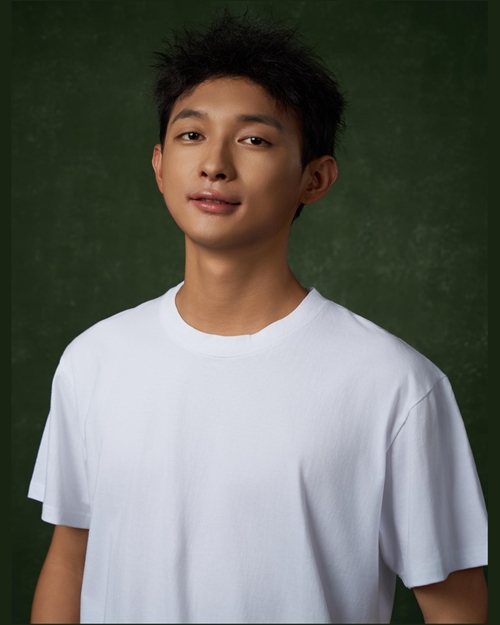}}]{Chenbo Shi}
  Chenbo Shi was born in 2000 in China. He received his Bachelor's degree from the University of Electronic Science and Technology of China (UESTC) in 2022. He is currently pursuing his Ph.D. at the same institution. His research interests include electromagnetic theory, characteristic mode theory, and computational electromagnetics. 
  
  Chenbo has been actively involved in several research projects and has contributed to publications in these areas. His work aims to advance the understanding and application of electromagnetic phenomena in various technological fields.
  \end{IEEEbiography}

  \begin{IEEEbiography}[{\includegraphics[width=1in,height=1.25in,clip,keepaspectratio]{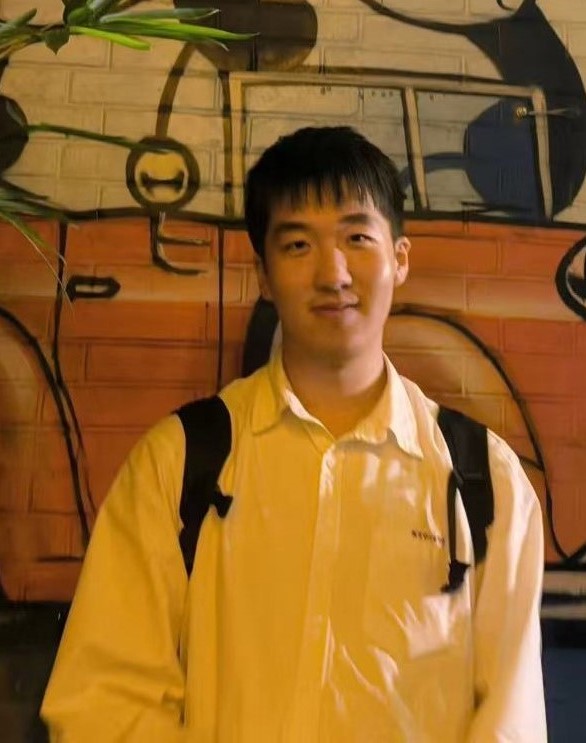}}]{Shichen Liang}
  received the B.E. degree from Beijing University of Chemical Technology (BUCT), Beijing, China, in 2022. He is currently pursuing the M.S. degree with the School of Electronic Science and Engineering, University of Electronic Science and Technology of China (UESTC), Chengdu, China. 
  
  His research interests include electromagnetic theory and electromagnetic measurement techniques.
  \end{IEEEbiography}

 \begin{IEEEbiography}[{\includegraphics[width=1in,height=1.25in,clip,keepaspectratio]{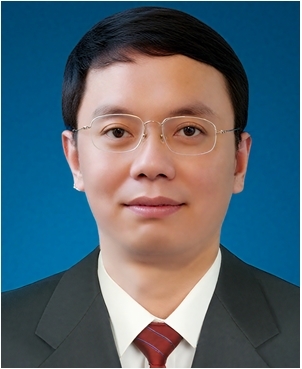}}]{Jin Pan}
    received the B.S. degree in electronics and communication engineering from the Radio Engineering Department, Sichuan University, Chengdu, China, in 1983, and the M.S. and Ph.D. degrees in electromagnetic field and microwave technique from the University of Electronic Science and Technology of China (UESTC), Chengdu, in 1983 and 1986, respectively. 
    
    From 2000 to 2001, he was a Visiting Scholar in electronics and communication engineering with the Radio Engineering Department, City University of Hong Kong. He is currently a Full Professor with the School of Electronic Engineering, UESTC. 
    
    His current research interests include electromagnetic theories and computations, antenna theories, and techniques, field and wave in inhomogeneous media, and microwave remote sensing theories and its applications. 
    \end{IEEEbiography}

  \begin{IEEEbiography}[{\includegraphics[width=1in,height=1.25in,clip,keepaspectratio]{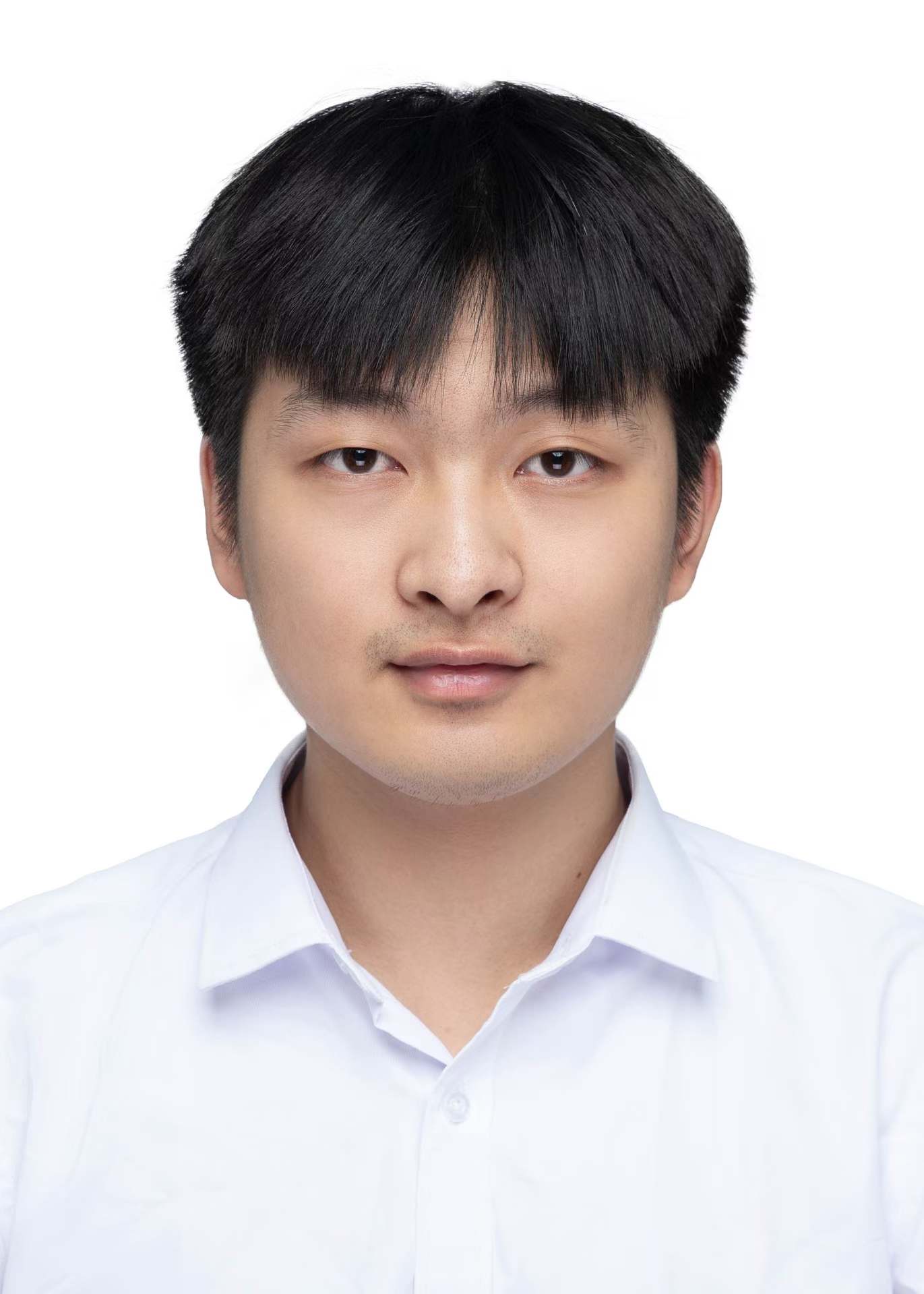}}]{Xin Gu}
  received the B.E. degree from Chongqing University of Posts and Telecommunications (CQUPT), ChongQing, China, in 2022. He is currently pursuing the M.S. degree with the School of Electronic Science and Engineering, University of Electronic Science and Technology of China (UESTC), Chengdu, China.
    
  His research interests include electromagnetic theory and electromagnetic measurement techniques
  \end{IEEEbiography}

  \begin{IEEEbiography}[{\includegraphics[width=1in,height=1.25in,clip,keepaspectratio]{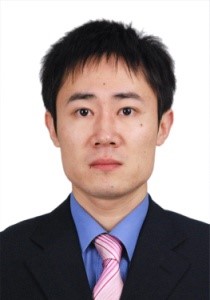}}]{Le Zuo}
  received the B.Eng., M.Eng. and Ph.D. degrees in electromagnetic field and microwave techniques from the University of Electronic Science and Technology of China (UESTC), in 2004, 2007 and 2018, respectively. 
  
  From 2017 to 2018, he was a Research Associate with the School of Electrical and Electronic Engineering, Nanyang Technological University, Singapore. He is currently a Research Fellow with the 29th Institute of the China Electronics Technology Group, Chengdu, China. His research interests include antenna theory and applications.
  \end{IEEEbiography}

\end{document}